\documentclass[10pt]{revtex4}
\usepackage{hyperref}
\usepackage{amsmath}
\usepackage[psamsfonts]{amssymb }
\usepackage{mathrsfs}
\begin{document}

\title{\Large Renormalizability of Supersymmetric Group Field Cosmology}

 \author{ Sudhaker Upadhyay}
 \email {sudhakerupadhyay@gmail.com; sudhaker@boson.bose.res.in}
 
\affiliation { S. N. Bose National Centre for Basic Sciences,\\
Block JD, Sector III, Salt Lake, Kolkata -700098, India. }

\begin{abstract}
 In this paper we consider the gauge invariant
  third quantized model of supersymmetric group field cosmology. The
supersymmetric BRST invariance for such theory in  non-linear gauge is also analysed.  The path integral formulation to the case of a multiverse made up of homogeneous
and isotropic spacetimes filled with a perfect
fluid
  is presented. The renormalizability  for the scattering of universes in multiverse are established with   suitably constructed master equations for connected diagrams and 
  proper vertices. The Slavnov-Taylor identities for this theory  hold to all orders of radiative corrections. 
\end{abstract}

\maketitle 
\section{Introduction}
In nonperturbative quantization of gravitational degrees of freedom  the background-independent loop quantum gravity 
 is widely investigated  \cite{rov,thi,thi1}. In this  quantization
scheme   the constraints are written in
terms of the densitized triad and of the Ashtekar-Barbero
connection \cite{as0, as, as1,as2,li,ge}. The main difficulties
of loop quantum gravity quantization scheme  are the lack of complete definition of
the quantum dynamics and the proof that leads back the resulting
theory  to Einstein’s gravity. The complete definition of the
quantum dynamics of spin network states has been obtained by embedding loop quantum gravity  states into the larger
framework of group field theories  \cite{ori1,ori2}  via
spin-foam models  \cite{ori,per}. These are basically quantum
field theories on group manifolds and the   Feynman amplitudes of   group field theories are spin-foam models. 
 The loop quantum gravity is a second quantized object with fixed topology. 
 However, the topology changing processes can not be analysed through second quantization approach
 and therefore we need the third quantization to analyse such processes \cite{bu,pi,pe,ma}. 
 The basic idea of the third quantization formalism is
 to treat the many-universe system as a quantum
field theory on superspace \cite{str}.
The name third quantization comes from
the fact that the field which is quantized is the wave function
of the universe, and it depends on the particles
existing within the universe.
 The third quantization of loop quantum gravity naturally leads to the group field theory \cite{ab, sm, ab1, ta}.
The minisuperspace (Wheeler-De Witt) approximations to  group field theory are known as group field cosmology  \cite{aa, qi, gu, ew, qi1, cal, gl,sg}.

Furthermore, the supersymmetry has been proved as an important ingredient in the study of M-theory \cite{il,lm,wt} which 
provides a basis for many phenomenological models beyond the standard model  
\cite{mg,mc}. The supersymmetry is also a prominent candidate for dark matter \cite{sa}. A supersymmetric
 generalization of group field cosmology, called as super-group field cosmology, has  been studied very recently
 as a model for multiverse 
 \cite{fai0}. However, a variety of multiverse hypotheses have been  
considered from different cosmological viewpoints \cite{bar}. The super-group field cosmology 
 remains invariant under gauge symmetry transformation and therefore contains some unphysical degrees of freedom. 
 The spurious  degree of freedom can be removed by choosing a suitable  gauge-fixing condition
 and this  can be achieved 
 by adding a term, called as gauge-fixing term, in the action at quantum level.
 This gauge-fixing term reflects the Faddeev-Popov ghosts in the void (the vacuum state of the multiverse)  functional (corresponds to vacuum
 functional  in the second quantization) to complete the
 effective theory.  The supersymmetric 
 BRST transformation as well as unitarity for the theory of multiverse has discussed very recently  \cite{fai}.
  So, it is worthwhile to explore
  the   identities between all 
  connected and  disconnected Green  functions to
  proof the algebraic renormalizability of multi-universe. This is the motivation of our
  present investigation. 
 
 In this paper we analyse the gauge invariance of super-group field cosmology
 which is a model for homogeneous
and isotropic multiverse  filled with a perfect
fluid.
 The linear and non-linear BRST invariance for such model of mutiverse are written explicitly.
 The Jacobian for such BRST transformation is unit.
 The symmetry generator   annihilates the physical states of total Hilbert space
 to prove the unitarity of the scattering matrix.  We construct the void functional 
 for third quantized cosmological model  in terms of  supersources. 
 The BRST invariance of the generating functional leads to the master equations for connected Green's functions and for proper vertices. The renormalizability of the theory is proved by checking the consistency
 condition of the the Salvnov-Taylor identities. 
  Here we notice that the Jacobian for path integral measure 
 under BRST transformation is unit.

The plan of the paper is as follows. In section II, we discuss the supersymmetric
group cosmology. The analyses of the third quantized  BRST transformations for 
super-group field cosmology are made in section III. Section IV is devoted to
establish the  renormalizability for the theory of multiverse.
The last section is reserved for discussions and conclusions. 
\section{Supersymmetric group field cosmology}
Let us consider a quantum multiverse made up of homogeneous
and isotropic universes filled up with a perfect
fluid. 
 Now, we start with
the loop quantum cosmology with a massless scalar field $\phi$ as matter in a spatially flat, homogeneous and
isotropic universe. The four-dimensional metric is then  given by
\begin{eqnarray}
ds^2=-N^2 (t)dt^2+a^2(t)\delta_{ab}dx^adx^b,
\end{eqnarray}
where $N(t)$ and $a(t)$ are lapse function and scaler factor respectively.  Here 
the spatial indices are labelled by Latin indices $a, b,...= 1, 2, 3$.
The variables used in loop quantum gravity are the   densitized
triad $E^a_i$ and the 
Ashtekar-Barbero  connection $A^i_a =\gamma (\omega^i _0)_a$, where $\gamma$
is the Barbero-Immirzi parameter and $(\omega^i _0)_a$
 is the spin connection compatible with the triad. The
curvature of $A^i_a$ in the loop quantum cosmology is expressed through the holonomy around a loop such that
the area of a loop   cannot be smaller than a fixed minimum area because
the smallest eigenvalue of the area operator in loop quantum gravity is non-zero.
Now one defines the    eigenstates of the volume operator $\cal{V}$  with a basis, $|\nu \rangle $,  are  
$ {\cal{V}} |\nu \rangle = 2 \pi \gamma G |\nu| |\nu \rangle
$, where gravitational conﬁguration variable $\nu = \pm a^2 {\cal{V}}_0 /2\pi \gamma G$ has the dimensions of length. 
 The Hamiltonian  constraint  for a homogeneous isotropic universe (in the Plank units) 
 is defined as \cite{cal}
\begin{equation} 
 - B(\nu)[E^2 - \partial^2_\phi] \Phi(\nu, \phi)= K ^2\Phi(\nu, \phi)  =0,\label{1}
\end{equation}
  where $\Phi(\nu, \phi)$ is a wave function on configuration space
  and $E^2 $ is a difference operator of the form:
   \begin{eqnarray}
- B(\nu) E^2 \Phi(\nu, \phi) =  C^+(\nu) \Phi(\nu+\nu_0 , \phi)
+C^0(\nu)\Phi(\nu, \phi)  +C^-(\nu)\Phi(\nu-\nu_0, \phi),
\end{eqnarray}
and $K_\mu$ is defined as   $K_{\mu} =\eta_{\mu\nu} K^\nu = (\sqrt{B(\nu)}\partial_\phi, \sqrt{B(\nu)} E)$   with metric $\eta_{\mu\nu}= \mbox{diag}(+1, -1)$.
So, $K ^2= K_\mu K^\mu =- B(\nu)[E^2 - \partial^2_\phi] $.
Here $\nu_0$ is an elementary length
unit, usually defined by the square root of the area gap
 and the functions $B(\nu), C^+(\nu),
C^0(\nu)$ and $C^-(\nu)$
  depend
on the choice of the lapse function and on  the details of quantization scheme.
For particular choice of lapse function, $N=1$, in an improved dynamic scheme, these 
functions have following form \cite{aste}:
\begin{eqnarray}
B(\nu)&=& \frac{3\sqrt{2}}{8\sqrt{\sqrt{3}\pi\gamma G}}|\nu | \left| \left|\nu +
\frac{\nu_0}{4}\right|^{\frac{1}{3}} -\left|\nu -
\frac{\nu_0}{4}\right|^{\frac{1}{3}}\right|^3,\nonumber\\
C^+(\nu)&=& \frac{1}{12\gamma\sqrt{2\sqrt{3}}}\left|\nu +\frac{\nu_0}{2}\right| \left| \left|\nu +
\frac{\nu_0}{4}\right|  -\left|\nu +
\frac{3\nu_0}{4}\right|\right|,\nonumber\\
C^0(\nu) &=& -C^+(\nu) -C^+(\nu -\nu_0),\nonumber\\
C^-(\nu)&=&  C^+(\nu-\nu_0).\label{fun1}
\end{eqnarray}
However, for $N=a^3$ where $a$ is scale factor and for orientation reversal symmetric wave function the structures of these functions are
\begin{eqnarray}
B(\nu)&=&\frac{1}{\nu},\nonumber\\
C^+(\nu)&=& \frac{\sqrt{3}}{8\gamma}\left(\nu +\frac{\nu_0}{2}\right),\nonumber\\
C^0(\nu) &=&  -\frac{\sqrt{3}}{4\gamma} \nu,\nonumber\\
C^-(\nu)&=& C^+(\nu-\nu_0),\label{fun}
\end{eqnarray}
and in the semiclassical limit $\nu\gg\nu_0$, these  expressions (\ref{fun})  agree 
with (\ref{fun1}).

By definition, the solutions of the first quantized theories correspond
to the free field solutions in
the second quantized formalism, the solutions of the second quantized theory should corresponds to free field solutions in the third quantized formalism. Hence, the solution of loop
quantum cosmology will now correspond to the classical field of group field cosmology. 
  Now, free field action  
for bosonic distribution of universes,
of which classical solution reproduces the
Hamiltonian constraint for loop quantum gravity, is given
by \cite{cal}
 \begin{equation}
 S_{b} =\sum_\nu \int d\phi\ {\cal L}_b= \sum_\nu \int d\phi \, \,    \Phi (\nu, \phi)  K^2  \Phi(\nu, \phi), 
\end{equation}
where  $\Phi(\nu, \phi)$ to be a real scalar
field.

It is worthwhile to  analyse the fermionic distribution of universes also which might
lead  
to correct the value of the cosmological constant. Since   the correct value
of cosmological constant is not obtained by considering  only bosonic distributions of universes in the multi-universe \cite{kolm}. Consequently, 
free action corresponding to fermionic  group field cosmology is constructed 
as \cite{fai0}
\begin{equation}
 S_{f} = \sum_\nu \int d\phi \, \,  \Psi^b(\nu, \phi) K_b^a \Psi_a(\nu, \phi),\label{fer}
\end{equation}
 where $\Psi_a (\nu, \phi) = (\Psi_1(\nu, \phi), \Psi_2(\nu, \phi) )$ is a fermionic spinor 
 field and $K_{ab}$ is defined as
  $K_{ab} = (\gamma^\mu )_{ab}K_\mu$. The spinor indices are raised
and lowered by the second-rank antisymmetric tensors $C^{ab}$ and $C_{ab}$, respectively.
These tensors satisfy following condition $C_{ab}C^{cb} = \delta^c_a$ \cite{van}.
Fermionic statics will not change the dynamics of a single universe by requirement
that the free action
(\ref{fer}) will lead to the Hamiltonian constraint   of loop quantum gravity.
The above bosonic and the fermionic actions describe bosonic and the fermionic universes
in the multiverse and hence, it is worthwhile to construct a supersymmetric gauge invariant
multiverse.
The  main idea behind the third quantization  is:
to  treat the many-universe system as a quantum
field theory on superspace.
 For this purpose we define, two complex scalar super-group fields
$\Omega(\nu, \phi, \theta) $ and $ \Omega^{\dagger}  (\nu, \phi, \theta)$ and a 
spinor super-group field $\Gamma_a (\nu, \phi, \theta)$, which are suitably contracted with generators 
 of a Lie algebra, $[T_A, T_B] = i f_{AB}^C T_C$, as
\begin{eqnarray}
   \Omega(\nu, \phi, \theta) &=&\Omega^A(\nu, \phi, \theta) T_A, \nonumber \\ 
\Omega^{\dagger}(\nu, \phi, \theta) &=& \Omega^{\dagger A}(\nu, \phi, \theta)  T_A \nonumber \\ 
\Gamma_a (\nu, \phi, \theta) &=& \Gamma_a^A (\nu, \phi, \theta)T_A. 
\end{eqnarray}
The extra variable $\theta$ are Grassmannian in nature which defines the extra direction in superspace.
The super-covariant derivative of these superfields is defined as 
\cite{fai0}
\begin{eqnarray}
  \nabla_a  \Omega^A(\nu, \phi, \theta)&=& D_a\Omega^A(\nu, \phi, \theta) -i f_{CB}^A\Gamma^C_a(\nu, \phi, \theta) \Omega^B(\nu, \phi, \theta),\nonumber\\
\nabla_a \Omega^{A \dagger}(\nu, \phi, \theta)  &=& D_a \Omega^{A \dagger}(\nu, \phi, \theta)  
+ i  f_{CB}^A\Omega^{C \dagger}(\nu, \phi, \theta)  \Gamma^B_a (\nu, \phi, \theta), 
\end{eqnarray}
where super-derivative $ D_a = \partial_a + K^b_a \theta_b$.
We also define the field strength for  a matrix valued spinor field ($ \Gamma^A_a $) as follows
$
 \omega^A_a (\nu, \phi, \theta) = \nabla^b \nabla_a \Gamma^A_b(\nu, \phi, \theta) $.
   With these
 introduction, now, we are  able to write the
 the classical action for the super-group field cosmology as (for details
 see e.g. \cite{fai0})
 \begin{eqnarray}
S_{0} =\sum_\nu \int d\phi \, \,    \left[D^2 \{\Omega_A^{\dagger} ( \nu, \phi, \theta) \nabla^2 \Omega^A
 ( \nu, \phi, \theta) 
  + \omega_A^a ( \nu, \phi, \theta) \omega^A_a( \nu, \phi, \theta)  \}\right]_|,
\end{eqnarray}
where ${}_|$ denotes $\theta =0$ after performing calculations.
This supersymmetric action  remains invariant under following gauge transformations
\begin{eqnarray}
  \delta \Omega^A(\nu, \phi, \theta) &=&  if_{CB}^A\Lambda^C (\nu, \phi, \theta)\Omega^B(\nu, \phi, \theta) ,\nonumber\\
\delta \Omega^{A \dagger}(\nu, \phi, \theta)  &=& -i f^A_{CB}\Omega^{C\dagger}(\nu, \phi, \theta) \Lambda^B(\nu, \phi, \theta), \nonumber\\
 \delta \Gamma^A_a(\nu, \phi, \theta) &=& \nabla_a \Lambda^A(\nu, \phi, \theta).
\end{eqnarray}
where $\Lambda^A$ is an infinitesimal local parameter. The gauge symmetry reflects
that theory posses  some redundant degrees of freedom. 
To quantize theory correctly we need to remove them.
In next section, we will show how these can be achieved for this theory. 
\section{The BRST symmetries and the physical states } 
In this section we discuss the nilpotent symmetries for the theory in linear \cite{fai} and in non-linear gauges. For this purpose, our first goal is to  remove the gauge redundancy by fixing a gauge. Making analogy with ordinary supersymmetric gauge theory, we chose the following covariant gauge condition:
\begin{equation}
 D^a \Gamma^A_a ( \nu, \phi, \theta) =0.
\end{equation}
This can be incorporated at a quantum level by adding the 
appropriate gauge-fixing term to  classical action which breaks the gauge symmetry. 
The linearised gauge-fixing term in Landau gauge using
Nakanishi-Lautrup auxiliary superfield  $  B^A( \nu, \phi, \theta) $ is given by
\begin{equation}
 S_{gf} = \sum_\nu \int d\phi \, \,   \left[D^2\{ B_A ( \nu, \phi, \theta) D^a \Gamma^A_a ( \nu, \phi, \theta)\}\right]_|.
\end{equation}
The effect of the  gauge-fixing term in the exponent of path integral
can be compensated by additional Faddeev-Popov ghost term. In this case
the ghost term is constructed as    
\begin{equation}
  S_{gh} = \sum_\nu \int d\phi \, \,   
 \left[D^2\{\bar c_A( \nu, \phi, \theta) D^a \nabla_a c^A
( \nu, \phi, \theta)\}\right]_|,
\end{equation}
where $   {c}^A( \nu, \phi, \theta)  $
and   $ \bar{c}^A ( \nu, \phi, \theta) $ are   
the ghost and anti-ghost superfields respectively.
Now the total action 
\begin{eqnarray}
S_T =\sum_\nu\int d\phi\ {\cal L}_T = S_{0}+  S_{gh}+  S_{gf},\label{act}
\end{eqnarray}   remains invariant under 
following    third quantized BRST transformations  
\begin{eqnarray}
  s\,\Omega^A(\nu, \phi, \theta)&= &if_{CB}
  ^Ac^C (\nu, \phi, \theta)\Omega^B(\nu, \phi, \theta) , \nonumber \\
s\,  \Omega^{A \dagger}(\nu, \phi, \theta)  &= & -i f^A_{CB}\Omega^{ \dagger C}(\nu, \phi, \theta)c^B(\nu, \phi, \theta), 
\nonumber \\ 
s\, c^A( \nu, \phi, \theta)&= &f^A_{CB}c^{  C}(\nu, \phi, \theta) c^B(\nu, \phi, \theta), 
\nonumber \\
 s\, \Gamma^A_a ( \nu, \phi, \theta) &= & \nabla_a c^A( \nu, \phi, \theta), 
\nonumber \\ 
s\, \bar{ c} ^A( \nu, \phi, \theta) &= & B^A( \nu, \phi, \theta),
 \nonumber \\
 s\, B^A( \nu, \phi, \theta) &= &0.\label{brs}
\end{eqnarray}
It is easy to check that this transformation is nilpotent in nature,   $s^2 =0$, and
therefore we are able to write the gauge-fixing and ghost terms collectively
in terms of BRST variation of gauge-fixed fermion  as \cite{fai}
\begin{eqnarray}
S_{gf}+S_{gh} =\sum_\nu\int d\phi \ s\left [D^2 \{\bar{c}_A  ( \nu, \phi, \theta) D^a \Gamma^A_a ( \nu, \phi, 
\theta) \}\right]_|.
\end{eqnarray}
The conserved charge corresponding to the BRST transformation
using Noether's theorem is calculated as \cite{fai}
\begin{eqnarray}
 Q_b  & = &   \sum_\nu \,\, 
\left[ \frac{ \partial \mathcal{L}_{T}  }{\partial \partial_{\phi} {\Gamma^A_a } (\nu, \phi, \theta )   } 
\nabla_a c^A( \nu, \phi, \theta)  +
 \frac{ \partial \mathcal{L}_{T}  }{\partial  \partial_{\phi} c^A (\nu, \phi, \theta )   } f^A_{CB}c^{  C}(\nu, \phi, \theta) c^B(\nu, \phi, \theta)
 \right.\nonumber \\ &&   +
\frac{ \partial \mathcal{L}_{T}  }{\partial  \partial_{\phi} \bar c^A (\nu, \phi, \theta )   }
  B^A( \nu, \phi, \theta) 
 + i
\frac{ \partial \mathcal{L}_{T}  }{\partial \partial_{\phi} \Omega^A (\nu, \phi, \theta )   }  f_{CB}
  ^Ac^C (\nu, \phi, \theta)\Omega^B(\nu, \phi, \theta) \nonumber \\ && \left.  -i
 \frac{ \partial \mathcal{L}_{T}  }{\partial \partial_{\phi} \Omega^{ \dagger A} (\nu, \phi, \theta )   } 
 f^A_{CB}\Omega^{ \dagger C}(\nu, \phi, \theta)c^B(\nu, \phi, \theta) \right]. 
\end{eqnarray}
The total vector superspace  of
the complete  theory (Eq. (\ref{act})) contains various unphysical states
as well as states with negative norm in addition to the physical states.
Consequently, the metric of this superspace and the inner product become
indefinite and a probabilistic description of the quantum theory is lost
unless we can restrict to a suitable super-subspace $(|\Psi \rangle )$ with a positive definite
inner product as
\begin{equation}
Q_b|\Psi \rangle =0. \label{ku}
\end{equation}
Furthermore, the  $\mathcal{S}$-matrix, which is BRST invariant, enable us to write  
\begin{equation}
[Q_b, \mathcal{S}]=0.
\end{equation}
If we define   an operator $\mathcal{S}_{phy}$ which acts and correspond to
the  $\mathcal{S}$-matrix in the physical super-subspace of states of the theory.
Then it must be unitary, i.e.
\begin{equation}
\mathcal{S}_{phy}^\dag \mathcal{S}_{phy}=\mathcal{S}_{phy}\mathcal{S}_{phy}^\dag =  {1}.
\end{equation}
The BRST invariance of the theory  automatically
leads to a formal proof of unitarity of the   super-$\mathcal{S}$-matrix in the super-subspace
of the truly physical states of the theory.

 Equation (\ref{ku}) gives us liberty to chose 
different gauge-fixing condition for the theory as physical states do not depend on the
choice of the gauge-fixing condition. 
The non-linear gauge conditions play important character
in second quantized field theories.
The investigation of non-linear gauge condition
in third quantized model of multiverse might play significant role.
For this purpose, we construct  
  the gauge-fixing and ghost terms in non-linear gauge condition for the theory 
of multiverse   as follows:
\begin{eqnarray}
S_{gf}+S_{gh} &= &\sum_\nu \int d\phi \, \,   
 \left[D^2\left\lbrace B_A ( \nu, \phi, \theta) D^a \Gamma^A_a (\nu, \phi, \theta)+\frac{1}{2}\bar c_A( \nu, \phi, \theta) D^a \nabla_a c^A
( \nu, \phi, \theta) \right.\right.\nonumber\\
&+&\left.\left.\frac{1}{8} f_{BC}^Af_{A}^{GH}\bar c^B (\nu, \phi, \theta)c^C(\nu, \phi, \theta) \bar c_G(\nu, \phi, \theta) c_H(\nu, \phi, \theta)\right\rbrace\right]_|.
\label{non}
\end{eqnarray}
The third quantize non-linear BRST transformations, under which the above expression
(\ref{non})
is invariant, are calculated as
\begin{eqnarray} 
  s\,\Omega^A(\nu, \phi, \theta)&= &if_{CB}
  ^Ac^C (\nu, \phi, \theta)\Omega^B(\nu, \phi, \theta) , \nonumber \\
s\,  \Omega^{A \dagger}(\nu, \phi, \theta)  &= & -i f^A_{CB}\Omega^{ \dagger C}(\nu, \phi, \theta)c^B(\nu, \phi, \theta), 
\nonumber \\ 
s\, c^A( \nu, \phi, \theta)&= &f^A_{CB}c^{  C}(\nu, \phi, \theta) c^B(\nu, \phi, \theta), 
\nonumber \\
 s\, \Gamma^A_a ( \nu, \phi, \theta) &= & \nabla_a c^A( \nu, \phi, \theta), 
\nonumber \\ 
s\, \bar{ c} ^A( \nu, \phi, \theta) &= & B^A( \nu, \phi, \theta) -\frac{1}{2}f^A_{BC} \bar c^B ( \nu, \phi, \theta) 
c^C(\nu, \phi, \theta),
 \nonumber \\
 s\, B^A( \nu, \phi, \theta) &= & -\frac{1}{2}f^A_{BC}  c^B ( \nu, \phi, \theta) 
B^C( \nu, \phi, \theta) \nonumber \\
& -& \frac{1}{8} f_{BC}^Af_{GH}^{C}c^B(\nu, \phi, \theta) c^G(\nu, \phi, \theta) \bar c^H
(\nu, \phi, \theta),\label{brs1}
\end{eqnarray}
which are nilpotent in nature, i.e. $s^2=0$.
\section{Renormalizability of the multiverse }
To study the quantum effects for third quantize group field cosmology
first we  define the source free void functional  
as
\begin{eqnarray}
\left\langle 0|0\right\rangle =Z[0]=\int {\cal D} M e^{iS_T},
\end{eqnarray}
where  ${\cal D} M\equiv {\cal D}\Omega{\cal D}\Omega^\dag{\cal D} \Gamma_a
{\cal D} B{\cal D}c{\cal D}\bar c$ is the path integral measure.
The above generating functional remains invariant under the infinitesimal BRST transformation
given in Eqs. (\ref{brs}) and (\ref{brs1}). It is easy to calculate the Jacobian for   such BRST transformations which
comes unit. 
Further to write full effective action
for the group field cosmology we need to add following external supersource term in the
$S_T$
\begin{eqnarray}
S_{ext}&=&\sum_\nu\int d\phi\ \left[\bar\eta_A(\nu,\phi, \theta )\Omega^A (\nu,\phi, \theta ) +\Omega_A^\dag (\nu,\phi, \theta )\eta^A(\nu,\phi, \theta )
+ J^a_A(\nu,\phi, \theta ) \Gamma_a^A(\nu,\phi, \theta )
\right.\nonumber\\
&+&\left.\bar \mu_A(\nu,\phi, \theta ) c^A (\nu,\phi, \theta )
+\bar c^A (\nu,\phi, \theta )\mu_A (\nu,\phi, \theta )
+ k^A (\nu,\phi, \theta ) B_A(\nu,\phi, \theta )\right. \nonumber\\
&+&\left.\bar\zeta_A (\nu,\phi, \theta ) s\Omega^A (\nu,\phi, \theta )   +  \zeta^A(\nu,\phi, \theta ) s\Omega_A^\dag(\nu,\phi, \theta )  
+K_a^A(\nu,\phi, \theta )  s\Gamma_A^a(\nu,\phi, \theta )\right. \nonumber\\
&+&\left.\xi_A (\nu,\phi, \theta )s c^A (\nu,\phi, \theta ) \right]_|,
\end{eqnarray}
where each the superfields are coupled with their external supersources. 
We have not only introduced supersources for all the field variables
in the theory, but we have also added supersources $(\bar \zeta_A, \zeta_A,
  K_a^A,  \xi_A  )$ for the
composite variations under the BRST transformation.

Now,  the generating functional for Green's functions, denoting all the supersources by $J$, is given by
\begin{eqnarray}
\left\langle 0|0\right\rangle^J = Z[J]=e^{iW[J]}=\int {\cal D} M e^{iS_{eff}},\label{gen}
\end{eqnarray}
where effective action is defined as $S_{eff}=S_T+S_{ext}$ and  $W[J]$ is the
as generating functional for only connected Feynman diagrams.

The void expectation values of superfields, in the presence of sources,
can now be written as
\begin{eqnarray}
\left\langle 0|\Omega^A (\nu,\phi, \theta )|0\right\rangle^J &=&\frac{\delta W[J]}{\delta \bar\eta_A
(\nu,\phi, \theta )},\ \ \
\left\langle 0|\Omega^{\dag A} (\nu,\phi, \theta )|0\right\rangle^J =\frac{\overleftarrow{\delta}  W[J]}{\delta  \eta_A
(\nu,\phi, \theta )},\nonumber\\
\left\langle 0|\Gamma_a^A (\nu,\phi, \theta )|0\right\rangle^J &=&\frac{\delta W[J]}{\delta J^a_A
(\nu,\phi, \theta )},\ \ \
\left\langle 0|c^A (\nu,\phi, \theta )|0\right\rangle^J =\frac{\delta W[J]}{\delta \bar\mu_A
(\nu,\phi, \theta )},\nonumber\\
\left\langle 0|  \bar c^A (\nu,\phi, \theta )|0\right\rangle^J &=&\frac{\overleftarrow{\delta} W[J]}{\delta \mu_A
(\nu,\phi, \theta )},\ \ \
\left\langle 0|  B^A (\nu,\phi, \theta )|0\right\rangle^J =\frac{\delta W[J]}{\delta k_A
(\nu,\phi, \theta )},\nonumber\\
\left\langle 0|  s\Omega^A (\nu,\phi, \theta )|0\right\rangle^J &=&\frac{\delta W[J]}{\delta \bar\zeta_A
(\nu,\phi, \theta )},\ \ \
\left\langle 0|  s\Omega^{\dag A} (\nu,\phi, \theta )|0\right\rangle^J =\frac{\delta W[J]}{\delta  \zeta_A
(\nu,\phi, \theta )},\nonumber\\
\left\langle 0|  s\Gamma_a^A (\nu,\phi, \theta )|0\right\rangle^J &=&\frac{\delta W[J]}{\delta K^a_A
(\nu,\phi, \theta )},\ \ \
\left\langle 0|  sc^A (\nu,\phi, \theta )|0\right\rangle^J =\frac{\delta W[J]}{\delta \xi_A
(\nu,\phi, \theta )}.
\end{eqnarray}
The invariance of generating functional for third quantized
super-group field cosmology given in Eq. (\ref{gen}) under BRST transformation
leads to
\begin{eqnarray}
&&\sum_\nu\int d\phi\ \left[\bar\eta_A(\nu,\phi, \theta )\frac{\delta W[J]}{\delta \bar\zeta_A
(\nu,\phi, \theta )}- \eta_A(\nu,\phi, \theta ) \frac{\delta W[J]}{\delta  \zeta_A
(\nu,\phi, \theta )} +J^a_A(\nu,\phi, \theta )\frac{\delta W[J]}{\delta J^a_A
(\nu,\phi, \theta )}\right.\nonumber\\
&&+\left.\bar\mu_A(\nu,\phi, \theta )\frac{\delta W[J]}{\delta  \xi_A
(\nu,\phi, \theta )}+\mu_A(\nu,\phi, \theta )\frac{\delta W[J]}{\delta k_A
(\nu,\phi, \theta )}  \right ]_|=0.\label{mas}
\end{eqnarray}
This is the master equation from which we can derive all the identities
relating the connected Green’s functions of the multiverse by taking
functional derivatives with respect to supersources.

Further, we construct the classical generating functional for proper (one particle irreducible) 
vertices, so-called vertex functional, using Legendre transformation as follows
\begin{eqnarray}
\Gamma^{(0)} [\Phi, J]=W[J]&-&\sum_\nu\int d\phi\ \left[\bar\eta_A(\nu,\phi, \theta )\Omega^A (\nu,\phi, \theta ) +\Omega_A^\dag (\nu,\phi, \theta )\eta^A(\nu,\phi, \theta )
\right.\nonumber\\
&+&\left.  J^a_A(\nu,\phi, \theta ) \Gamma_a^A(\nu,\phi, \theta )+\bar \mu_A(\nu,\phi, \theta ) c^A (\nu,\phi, \theta )
\right.\nonumber\\
&+&\left. \bar c^A (\nu,\phi, \theta )\mu_A (\nu,\phi, \theta ) + k^A (\nu,\phi, \theta ) B_A(\nu,\phi, \theta )\right ]_|,
\end{eqnarray} 
where $\Phi$ and $J$ are the generic  notation for superfields and supersources respectively.

Now,  the master equation (Slavnov-Taylor identity) for proper vertices
of Feynman diagram of universes has the following form: 
\begin{eqnarray}
{\cal S}(\Gamma^{(0)} )=\sum_\nu\int d\phi && \left[\frac{\overleftarrow{\delta} \Gamma^{(0)} [\Phi, J]}{\delta \Omega^A
(\nu,\phi, \theta )}\frac{\delta \Gamma^{(0)} [\Phi, J]}{\delta \bar\zeta_A
(\nu,\phi, \theta )}+\frac{\delta\Gamma^{(0)} [\Phi, J]}{\delta  \zeta_A
(\nu,\phi, \theta )} \frac{\delta \Gamma^{(0)} [\Phi, J]}{\delta \Omega^{\dag A}
(\nu,\phi, \theta )} \right.\nonumber\\
 &&+ \left. \frac{\delta \Gamma^{(0)} [\Phi, J]}{\delta \Gamma_a^A
(\nu,\phi, \theta )}\frac{\delta \Gamma^{(0)} [\Phi, J]}{\delta J^a_A
(\nu,\phi, \theta )}+\frac{\overleftarrow{\delta} \Gamma^{(0)} [\Phi, J]}{\delta c^A
(\nu,\phi, \theta )}\frac{\delta \Gamma^{(0)} [\Phi, J]}{\delta  \xi_A
(\nu,\phi, \theta )} \right.\nonumber\\
&&+ \left.  B_A(\nu,\phi, \theta )\frac{\delta \Gamma^{(0)} [\Phi, J]}{\delta \bar c^A
(\nu,\phi, \theta )}    \right ]_|=0. 
\end{eqnarray}
This relationship is essential in proving the renormalizability
of the super-gauge group theories. Consequently,
there is no direct obstruction in calculating the Feynman diagram
for multi-scattering precesses. 
This provides positive signature
to calculate renormalizable scattering of universes  in the
multiverse. 
\section{The consistency check}
To extend the Slavnov-Taylor identity to all orders of perturbation theory we
consider the 
vertex functional written as a power series in $\hbar$ 
\begin{equation}
 \Gamma  [\Phi, J] =\sum_{n=0}^\infty \hbar^n \Gamma^{(n)} [\Phi, J].
\end{equation}
The
Slavnov-Taylor identity for the full vertex functional defined as
\begin{eqnarray}
{\cal S}(\Gamma )=0. 
\end{eqnarray}
Similar to  tree level  approximation, we  define the theory by gauge-fixing condition as,
\begin{eqnarray}
\frac{\delta \Gamma[\Phi, J] }{\delta B^A}=D^2 D^a\Gamma_a^A.\label{rad}
\end{eqnarray}
To show the stability of the above expression to all order of radiative corrections, we assume that
the relation (\ref{rad}) is valid upto $(n-1)^{th}$ order in $\hbar$.
We write the most general breaking term compatible with power-counting
as
\begin{equation}
\frac{\delta \Gamma[\Phi, J] }{\delta B^A}=D^2 D^a\Gamma_a^A +\hbar^n \Delta^A +O(\hbar^{n+1}),
\end{equation}
with the breaking
\begin{equation}
\Delta^A (\nu, \phi,\theta) =D^2H^A(\Gamma_a^A, \bar c^A, c^A),
\end{equation}
where $H^A$ is some local polynomial in $\Gamma_a^A, \bar c^A, c^A$.
In this case the $ \Delta$ satisfies following consistency condition  
\begin{eqnarray}
\frac{\delta}{\delta  B^A (\nu, \phi, \theta)} \Delta^B(\nu, \varphi, \theta)  -\frac{\delta}{\delta B^B (\nu, \varphi, \theta)} \Delta^A (\nu, \phi, \theta)=0.
\end{eqnarray}
 Following from 
 $[\delta/\delta B^A (\nu, \phi, \theta), \delta/\delta B^B(\nu, \varphi, \theta)]= 0$, the above condition (which is an integrability condition) shows that the breaking
can be written as a functional derivative of $B^A(\nu,\phi,\theta)$ as
\begin{equation}
\Delta^A (\nu, \phi, \theta) =\frac{\delta}{\delta B^A (\nu, \phi, \theta)}\tilde\Delta,\ \mbox{with}\ 
  \tilde\Delta =\sum_\nu\int d\varphi\ D^2\left(B_A H^A (\Gamma_a^A, \bar c^A, c^A)\right)(\nu,\varphi,\theta).
  \end{equation}
 The absorption of $-\tilde\Delta$ as a counter term  at the order $n$ assures the validity of Eq. (\ref{rad})
up to order $n$. This shows that the gauge condition is consistent to all orders of 
 perturbative super-group field cosmology which ends the recursive proof of 
 renormalizability of the theory. One can also check the consistency with anti-ghost equation which must hold to
 all order of perturbation. 
\section{Conclusion}
In this paper, we have considered the theory of  super-group field
cosmology which is a model for multiverse. In multiverse scenario, it is considered that   the universes can collide with each other to form the other
universes. In this consideration, the big bang and creation of the universe  is 
 a nothing more than the collision between multiverse. It is also possible that even our own universe
formed because of the collision of two previous universes.
It has been found that the third quantized model for multiverse is gauge invariant.
Further, the third quantized linear and non-linear BRST transformations and corresponding generators have been investigated. The unitarity
of scattering matrix  for the physical processes in multiverse has been proved.
 We have analysed the quantum effects for  super-group field cosmology through path integral approach.
The requirement of renormalizabilty has been fulfilled by 
deriving the Slavnov-Taylor identity for the theory of cosmology.  
The master equations for this theory have also been derived from which  all the identities
relating the connected Green’s functions  and relations between various  proper vertices 
can be established. The validity oft Slavnov-Taylor identity  to all order of perturbation through a consistency check assures, or atleast there is
no direct obstructions in, the algebraic proof of
renormalizabilty of multiverse. Furthermore, we have noticed that the Jacobian  for the path integral measure under BRST transformations
are unit. It will be interesting to generalize the third quantized BRST transformation for this model
which will lead some non-trivial Jacobian for path integral measure \cite{sud}.
Since generalized BRST transformations have found many applications in the second quantized gauge field theories 
\cite{sb,sdj,rb,smm,fs,sud1,susk, subp1,ssb,sudha,rbs}. 
The  computation of
corrections in  the theory of multiverse with
Batalin-Vilkovisky approach will be  brilliant \cite{sudhaker}.

\end{document}